\documentclass[conference]{IEEEtran}
\usepackage[T1]{fontenc}
\IEEEoverridecommandlockouts
\usepackage{cite}
\usepackage{amsmath,amssymb,amsfonts}
\usepackage{algorithmic}
\usepackage{graphicx}
\usepackage{textcomp}
\usepackage{booktabs}     
\usepackage{stfloats}
\usepackage{float}
\def\BibTeX{{\rm B\kern-.05em{\sc i\kern-.025em b}\kern-.08em
    T\kern-.1667em\lower.7ex\hbox{E}\kern-.125emX}}

\newcommand \ignore[1]{}

\usepackage{url}
\usepackage{hyperref}
\usepackage[table,x11names]{xcolor}
\usepackage{caption}
\usepackage{subcaption}

\usepackage{comment}
\definecolor{LightCyan}{rgb}{0.88,1,1}
\definecolor{pink}{rgb}{1,0,1} 

\makeatletter
\newcommand{\printfnsymbol}[1]{%
  \textsuperscript{\@fnsymbol{#1}}%
}
\makeatother

\begin{document}

\title{Mapping Multimodal Pilot Stress and Fatigue During Flight Sessions
}

\author{
Atandrila Chowdhury$^{1}$, Sudip Vhaduri$^{2}$, Julius Keller$^{3}$, Debra Henneberry$^{3}$, and Mark Wilson$^{4}$\\

$^{1}$Edwardson School of Industrial Engineering,
$^{3}$School of Aviation and Transportation Technology,\\
$^{4}$School of Human and Health Sciences,
Purdue University, IN 47907\\
$^{2}$Department of Computer Science,
University of Alabama, AL 35487\\
}


\maketitle
\textcolor{black}{\begin{abstract}
This study analyzes patterns of stress and exhaustion among student pilots throughout flight training using a combination of physiological and self-reported measurements. The Perceived Stress Scale (PSS-10) was used to measure perceived stress and exhaustion before and after each flight, while physiological data, including heart rate (HR), electrodermal activity (EDA), skin temperature, and acceleration, were continuously recorded during flight sessions. To identify recurring patterns in arousal and workload, physiological signals were preprocessed and analyzed across the flight stages. The findings indicate a buildup of workload-related weariness over time, as evidenced by steady increases in EDA and skin temperature across flights, as well as post-flight increases in self-reported exhaustion. Heart rate responses were more event-specific, with brief spikes during high-demand phases of flight. Overall, the findings demonstrate the value of combining physiological signals with subjective reports to identify patterns of stress and fatigue during real-world flight training and highlight the potential of data-driven approaches for monitoring pilot well-being.
\end{abstract}
}

\section {Introduction}
Flight operations are demanding environments where cognitive and physiological stress can accumulate over time. Chronic exposure can lead to fatigue, reduced alertness, and impaired performance. This paper aims to explore the interaction between physiological stress markers and subjective sleep quality among pilots.
\subsection{Motivation}
Flight operations demand constant attention, decision-making, and emotional control -  conditions that make stress and fatigue unavoidable components of a pilot’s experience. Stress and fatigue can significantly reduce pilot alertness, cognitive functioning, judgment, decision-making, memory, and attention. These impairments increase the likelihood of memory lapses, operational mistakes, poor decisions, and other safety risks, and in severe cases, may even result in unintended drowsiness during flight. 

Fatigue has been repeatedly identified as a contributing factor in accidents~\cite{goode2003pilots}. Data from NASA’s Aviation Safety Reporting System indicate that approximately 52,000 incidents- around 21\% of all reports – have been explicitly attributed to fatigue~\cite{mccallum2003commercial}.  Furthermore, a joint statement by 28 leading sleep scientists identified fatigue as the most significant and preventable cause of transportation–related accidents, accounting for an estimated 15-20\% of all such events~\cite{akerstedt2000consensus}.

Previous studies have shown that pilots’ psychophysiological parameters provide valuable insight into their mental and physical states~\cite{wilson1990use,bonner2002heart,hoogeboom2004does,haarmann2009combining,johannes2008individualized}. Common stressors identified by flight personnel include in-flight workload (20.8\%), poor communication quality (11.1\%), uncooperative passengers (18.1\%), and night flying (5.6\%)~\cite{socha2018flying}. A pilot’s psychological condition during various flight tasks is a critical factor influencing flight comfort, precision, and overall safety~\cite{veltman1996physiological}. Exposure to such conditions increases the pilot’s psychological burden, creating situations that require effective stress management to maintain performance and flight safety.

Stress and fatigue are key factors that directly affect how safely and efficiently pilots perform. Even with modern technology and automation, the human element remains the most critical part of flight operations. Small drops in focus or alertness can lead to serious mistakes. Understanding and monitoring pilot stress and fatigue are therefore essential for improving performance, reducing risk, and keeping flight operations safe.

\subsection{Related work}
Prior studies have looked at the physiological and psychological demands on pilots, emphasizing how tiredness and stress can affect flying performance and safety~\cite{gomez2024assessing}. Pilots frequently experience high levels of stress due to their workload, erratic schedules, and extended duty periods, according to numerous studies~\cite{socha2018flying}. These stressors have been associated with shorter reaction times, decreased attention, and poor decision-making, all of which raise the possibility of human error in aviation~\cite{rowden2011relative}. Prior studies have also highlighted the importance of identifying risk factors for fatigue, noting that these disorders can be avoided if identified early. Therefore, enhancing flight safety and performance outcomes requires an understanding of the cognitive, physiological, or environmental mechanisms that cause pilot fatigue~\cite{jun2023pilot}.

The NASA Task Load Index (NASA-TLX), the Multidimensional Fatigue Inventory (MFI), and other surveys have historically been used in research on pilot workload~\cite{gomez2024assessing}. Recent research has extended to objective monitoring using physiological signals such as heart rate variability (HRV), electroencephalography (EEG), and galvanic skin response (GSR), even though these tools capture subjective perceptions~\cite{regula2014study, setz2009discriminating, luzzani2024eda}. These indicators enable researchers to measure stress and fatigue responses across various flight phases and provide continuous, real-time insights into a pilot's psychophysiological state. The strong connection between task intensity and fatigue development has been confirmed by studies using flight simulators, which show that physiological stress markers rise dramatically during high-demand maneuvers and emergency situations. Despite these developments, comprehensive methods that incorporate physiological and self-reported data are still required to gain a deeper understanding of how pilots perceive and cope with stress in practical operational environments.

\subsection{Contributions}
In this study, we use physiological and self-reported data gathered during flight sessions to examine pilot stress and tiredness. Understanding how flight conditions and individual habits affect pilots' mental and physical states is the main objective. We integrate objective indicators such as heart rate, electrodermal activity, and body temperature with subjective assessments collected through pre- and post-flight surveys and sleep questionnaires to discover consistent patterns of stress and weariness. Deeper understanding of how pilots react to flight situations is made possible by this integrated approach. Our results can guide future approaches to fatigue management and safety enhancement in aviation training environments, as well as aid in the creation of data-driven models for tracking pilot well-being.

\section{Data Collection And Methods}


\subsection{Setting and Participants}
Four healthy student pilots who were enrolled in Purdue University's flight training program participated in this study. At the time of data collection, they were all in the early stages of their aviation careers, had logged between 100 and 300 flying hours, and had no self-reported history of neurological, cardiovascular, or sleep issues. Participants ranged in age from 19 to 20 (three men and one woman). 

Each participant had a different flight profile. Cross-country training flights were carried out by pilots Alpha, Bravo, and Delta. These flights usually comprised constant cruising phases at higher altitudes with little task change and maneuvering. Pilot Charlie, on the other hand, finished a local training flight that concentrated on emergency procedure simulations and performance maneuvers, including many engine-failure drills at lower altitudes in warmer weather. When assessing physiological responses among individuals, these variations in flight type and environmental exposure were taken into account.

Prior to data collection, each participant gave written informed consent, and participation was entirely voluntary. Each subject wore physiological monitoring sensors to capture skin temperature, heart rate (HR), electrodermal activity (EDA), and acceleration (ACC) during supervised flight sessions. 
In addition, they completed pre- and post-flight surveys, and responses to the Perceived Stress Scale (PSS) and the Karolinska Sleep Questionnaire (KSQ) were also taken into account. The study protocol was approved by the Institutional Review Board (IRB).

\subsection{Preliminaries And Data Collection}

\subsubsection{Physiological Measures}
A wearable device with several sensors on the wrist was used to continuously capture physiological information throughout each flight. Heart rate (HR), electrodermal activity (EDA), temperature (TEMP), and accelerometer (ACC) data were all simultaneously recorded by the device.  
While acceleration was utilized to identify motion-related artifacts, heart rate, EDA, and temperature were chosen as markers of physiological arousal.

\subsubsection{Sleep Quality and Non-Restorative Sleep}
The Karolinska Sleep Questionnaire (KSQ), which evaluates several aspects of sleep disturbance and recuperation, provided sleep-related variables.
The KSQ consists of three key factors: \textit{Sleep Quality}, \textit{Non-Restorative Sleep (NRS)}, and \textit{Sleep Apnea}~\cite{nordin2013psychometric}.

\begin{table}[htbp]
\centering
\caption{Factor loadings by sleep-related variables from the Karolinska Sleep Questionnaire (KSQ)}
\label{tab:ksq_short}
\renewcommand{\arraystretch}{1.2}
\resizebox{\columnwidth}{!}{%
\begin{tabular}{p{6cm}ccc}
\toprule
\textbf{Question} & \textbf{Sleep Quality} & \textbf{Non-Restorative Sleep} & \textbf{Sleep Apnea} \\
\midrule
Difficulties falling asleep &  \textbf{0.67} & 0.20 & 0.02 \\
Repeated awakenings with difficulties falling asleep again & \textbf{0.85} & -0.03 & 0.10\\
Premature awakenings & \textbf{0.78} & 0.05 & 0.12 \\
Disturbed/restless sleep & \textbf{0.81} & 0.20 & 0.09\\
Difficulties waking up &  -0.05 & \textbf{0.89} & 0.05 \\
Not well-rested on awakening &  0.53 & \textbf{0.63} & 0.08 \\
Felling of being exhausted at awakening &  0.53 & \textbf{0.63} & 0.16 \\
Heavy snoring &  0.08 & 0.01 & \textbf{0.82}\\
Gasping for breath during sleep &  0.11 & 0.09 & \textbf{0.89}\\
Cessation of breathing during sleep &  0.11 & 0.09 & \textbf{0.89}\\
\bottomrule
\end{tabular}
} 
\\[2pt]
\footnotesize{\textit{Note:} Bolded values indicate the strongest factor loadings for each question.}
\end{table}

In the present study, only the first two factors—Sleep Quality and NRS—were analyzed, as they are most relevant to stress and fatigue outcomes~\cite{tafoya2023sleep,tinajero2018nonrestorative}. 

\begin{figure*}[hbt!]
    \centering
    \begin{subfigure}[b]{0.95\linewidth}
        \includegraphics[width=\linewidth]{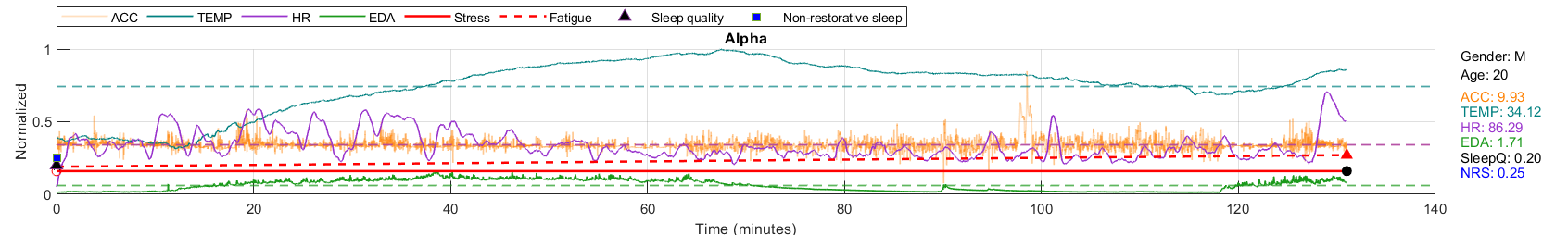}
        \caption{Physiological data – Pilot Alpha}
    \end{subfigure}
    \begin{subfigure}[b]{0.95\linewidth}
        \includegraphics[width=\linewidth]{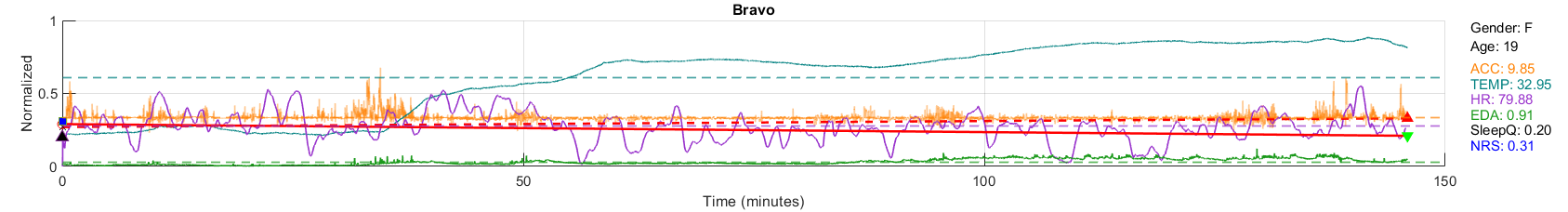}
        \caption{Physiological data – Pilot Bravo}
    \end{subfigure}
    \begin{subfigure}[b]{0.95\linewidth}
        \includegraphics[width=\linewidth]{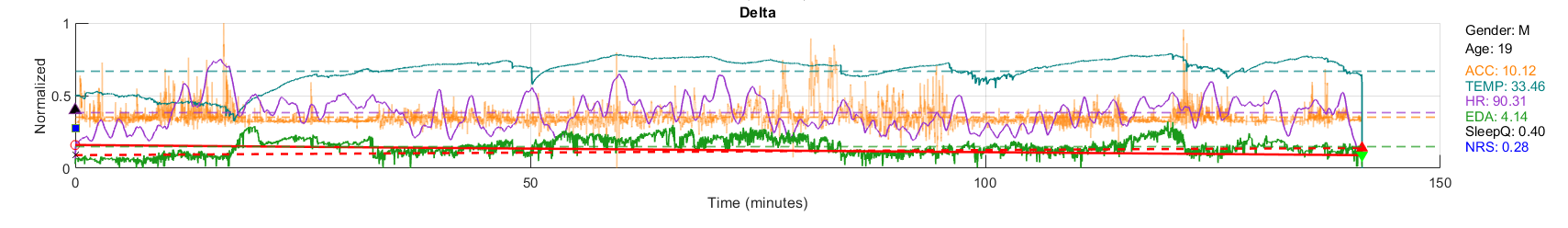}
        \caption{Physiological data – Pilot Delta}
    \end{subfigure}
    \begin{subfigure}[b]{0.95\linewidth}
        \includegraphics[width=\linewidth]{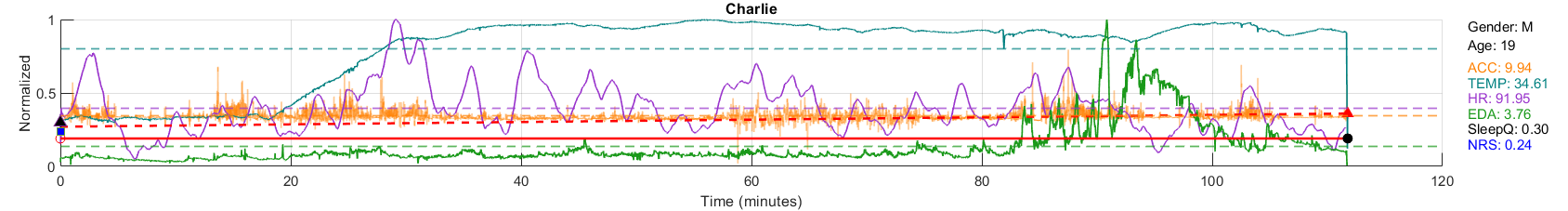}
        \caption{Physiological data – Pilot Charlie}
    \end{subfigure}
    \caption{Combined physiological data across all pilots (Temperature, EDA, HR, and ACC).}
    \label{fig:alldataimage}
\end{figure*}

The \textbf{Sleep Quality} subscale included four items~\cite{nordin2013psychometric}:
\begin{itemize}
    \item Difficulties falling asleep
    \item Repeated awakenings (with difficulties)
    \item Premature (final) awakening
    \item Disturbed or restless sleep
\end{itemize}
These items evaluated sleep continuity and disturbance, representing stress-related sleep problems~\cite{tafoya2023sleep}.

The \textbf{Non-Restorative Sleep (NRS)} subscale included ten items~\cite{nordin2013psychometric}:
\begin{itemize}
    \item Difficulties awakening
    \item Insufficient sleep
    \item Not feeling refreshed at wake-up
    \item Feeling of exhaustion at awakening
    \item Sleepiness during work
    \item Sleepiness during free time
    \item Mental fatigue
    \item Unintentional dozing off (naps) during work
    \item Unintentional dozing off (naps) during free time
    \item Fighting sleep to stay awake
\end{itemize}
These items reflected restorative sleep and daytime functioning, representing fatigue-related recovery~\cite{tinajero2018nonrestorative}.

All items were rated on a six-point scale ranging from 0 (“never”) to 5 (“always”).  
Each subscale score was computed as the mean of its item responses, and then normalized by dividing by the maximum possible score (5).  
Higher Sleep Quality scores indicate greater sleep disturbance (stress-related), whereas higher NRS scores indicate poorer restorative sleep (fatigue-related).
\\

\subsubsection{Self-Reported Stress and Fatigue}
Perceived stress and fatigue were assessed using the 10-item Perceived Stress Scale (PSS-10) administered in a Visual Analog Scale (VAS) format~\cite{smith2014assessment}. The PSS-10 was used to measure background levels of perceived stress.

The VAS format was also used to assess flight-related stress and fatigue immediately before and after each flight. For each construct, participants were presented with a horizontal line anchored at the left by “no stress” or “no fatigue” and at the right by “extreme stress” or “extreme fatigue.” Participants indicated their perceived level by drawing a vertical mark along the line, providing a continuous measure of subjective intensity.

Participants also rated the occurrence of unexpected events during the flight, the perceived difficulty of the task, and their overall flight performance. VAS responses were digitized, converted to numerical values ranging from 0 to 1, and used to compute pre–post differences in stress and fatigue for each participant.

\subsection{Pre-processing}
To maintain signal consistency among pilots, physiological and survey data were cleaned before analysis. Each session's raw recordings were loaded and formatted appropriately. Since the signal was recorded across x, y, and z for the accelerometer (ACC), the three raw axes were maintained in their original device units. The magnitude was calculated by taking the square root of the sum of the squared values from all three axes. Temperature (TEMP), heart rate (HR), and electrodermal activity (EDA) data were collected from their respective files, with zeros, NaNs, and other invalid samples eliminated. Before analysis, all signals, including the ACC magnitude, were standardized by taking the minimum and maximum values for each individual.

\section{Data Analysis and Visualization}
To visualize the entire dataset across subjects, all physiological parameters (HR, EDA, TEMP, and ACC) were plotted simultaneously. This representation, shown in Fig.~\ref{fig:alldataimage}, helps identify discrete activity zones that correlate with flying periods and provides a clear summary of how each signal behaves over time. Key areas of interest, such as flight commencement or landing, were identified for in-depth investigation using these maps. This procedure ensured that future analyses focused on the most relevant time frames that capture genuine physiological and behavioral changes during flight. Each pilot's unique physiological profile was thoroughly analyzed in order to better understand these collective patterns. This allowed for a closer look at signal variability, intra-flight dynamics, and pilot-specific stress reactions.

\subsection{Unimodal Analysis}
\label{sec:unimodal}

\subsubsection{Temperature}
To better separate physiological changes associated with flight activity, wrist temperature data were averaged over different time frames throughout the analysis. Each pilot's mean wrist temperature during the 20 minutes before the flight (0–20 min) was compared to the mean temperature during the active flight window (about 20 minutes after the device was worn until around 35 ± 5 minutes before its removal). Flight logs confirmed that this time frame nearly matched the actual engine start-to-shutdown timeframe. The obtained average values depict each pilot’s thermal profile at rest and in flight, allowing comparison of their respective internal temperature changes under similar ambient and workload conditions. 
All four pilots demonstrated a positive $\Delta$ Temp (ranging from
+1.97~\textdegree C  to +4.80 ~\textdegree C), indicating a
consistent increase in wrist temperature during the pre-flight baseline.

\begin{table}[H]
\caption{Average wrist temperature comparison across pre-flight and in-flight phases}
\begin{center}
\resizebox{\columnwidth}{!}{%
\begin{tabular}{c|c|c|c}
\hline
\textbf{Pilot} & \textbf{Avg. Temp (0–20 min)} & \textbf{Avg. Temp (20–$t_{max}$–40)} & \boldmath$\Delta$ \textbf{Temp (°C)} \\
\hline
Alpha   & 31.041 & 34.856 & +3.815 \\
Bravo   & 29.836 & 32.712 & +2.877 \\
Charlie & 30.587 & 35.390 & +4.803 \\
Delta   & 31.706 & 33.678 & +1.972 \\
\hline
\end{tabular}
}
\label{tab:pilotTempChange}
\end{center}
\end{table}

The pattern among participants demonstrates that temperature constantly climbed during flight, while the size of that increase varied. Pilot Charlie demonstrated the largest shift, which lines up with the unique parameters of that flight: lower altitude, greater ambient temperatures, and a more challenging maneuver profile. The remaining pilots displayed less pronounced but still noticeable increases in wrist temperature during flight, indicating that environmental factors, aircraft workload, and individual physiological reactions all interact together to form wrist temperature rather than any one factor alone.

\subsubsection{Electro dermal activity}
Electrodermal activity (EDA) was assessed to measure sympathetic arousal patterns during the flight duration for each subject. Data from Pilot Alpha were omitted due to signal dropout midway through the session, characterized by a flat line near zero, which may have been caused by a lack of skin contact with the electrodes. All three of the remaining pilots, i.e., Bravo, Charlie, and Delta, showed a distinct rising trend in EDA over time, suggesting elevated physiological arousal linked to flying activity.

From a pre-flight mean of 0.304 µS to 1.540 µS post-flight, Pilot Bravo demonstrated a little but consistent rise (slope = 0.0109 µS/min, $\Delta$EDA = 1.236 µS). With prominent peaks that corresponded to high-workload phases like simulated engine breakdowns, Pilot Charlie showed the highest response (slope = 0.0654 µS/min, $\Delta$EDA = 5.276 µS). Pilot Delta likewise demonstrated an increasing trend (slope = 0.0056 µS/min, $\Delta$EDA = 1.561 µS), suggesting significant arousal throughout the flight. All valid subjects showed positive EDA slopes overall, indicating a steady rise in sympathetic activity during and after the flight, consistent with the anticipated physiological response to environmental and cognitive stressors.

\begin{figure}[H]
    \centering
    \includegraphics[width=\columnwidth]{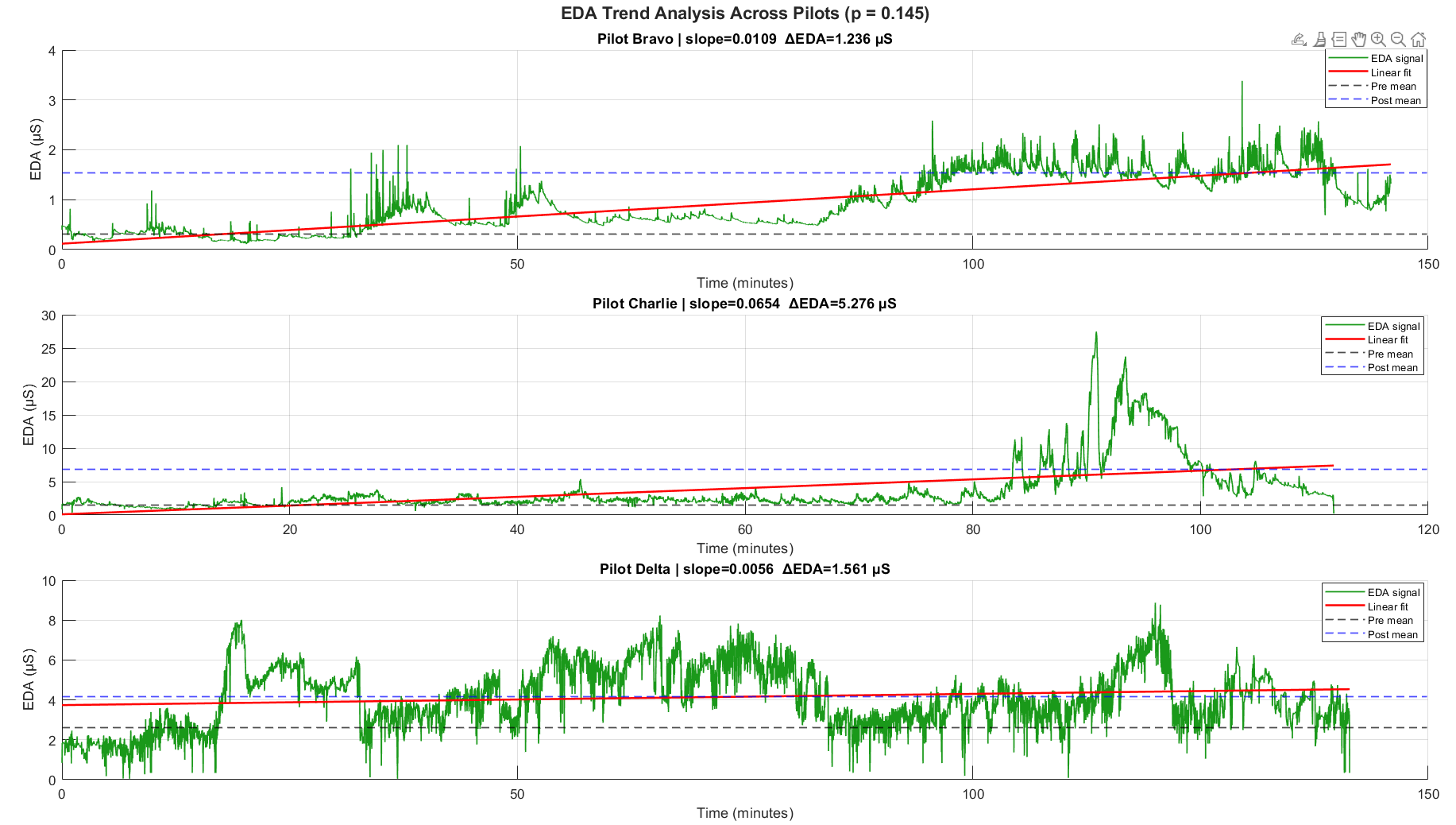}
    \caption{Electrodermal activity trends across pilots (Bravo, Charlie, Delta).}
    \label{fig:EDA}
\end{figure}

Environmental factors, such as elevated cabin temperature, may have contributed to this rise because EDA is sensitive to both heat and skin conductivity conditions. But EDA is also closely associated with sympathetic activation and psychological arousal, suggesting that the observed increase likely reflects the challenge or unpleasantness of each flight.
Notably, for Pilots Charlie and Delta, considerable EDA peaks were seen soon after the missions finished. This post-flight surge in both scenarios can be attributed to physical strain since each pilot had to manually drive the aircraft back to its parking spot. Similar acceleration (ACC) peaks recorded throughout the same time period support this theory by demonstrating greater physical activity in line with the aircraft control task.


\subsubsection {Accelerometer data}
Based on the ACC signal, both Charlie and Delta show a clear rise in movement for roughly 5–8 minutes after the end of the flight. This sustained increase stands out from their mid-flight baseline and is consistent with the continuous physical effort required when manually pushing the aircraft back into position. In contrast, Alpha and Bravo do not exhibit any comparable post-flight increase in ACC activity, indicating that they were not engaged in the same kind of prolonged physical movement after landing.
\subsubsection{Heart Rate (HR) Baseline Analysis}

\leavevmode\\
\noindent{\bf Pre-flight:} Pilot Alpha, Bravo, Delta show more time above the baseline (positive values), suggesting elevated heart rates before takeoff, which could indicate stress or anticipation. Pilot Charlie shows more time below the baseline (negative value), which might suggest a relaxed or calm state compared to the flight.

\noindent{\bf Takeoff:} Pilot Charlie and Pilot Delta show significant increases in HR (positive values) during takeoff, which is typical as the body responds to the stress or excitement of starting the flight. Pilot Bravo Shows a negative HR value during takeoff, suggesting that their heart rate was below baseline at this phase. Pilot Alpha shows a slight positive value, indicating that their HR might have been more stable.

\noindent{\bf Landing:}  Pilot Alpha and Pilot Charlie show a significant decrease in HR during landing (close to -1), indicating more time below the baseline, possibly due to relaxation or the lower physical demands of the landing phase. Pilot Delta has a less extreme negative value, which could indicate a slight decrease in HR during landing, but not as drastic. Pilot Bravo has a small positive value, which could imply that the pilot was in a relaxed or calm state during the flight.

\begin{table}
\caption{Heart Rate (HR) Time Above - Below Baseline for Each Pilot Across Flight Phases}
\begin{center}
\resizebox{\columnwidth}{!}{%
\begin{tabular}{|l|c|c|c|c|c|}
\hline
\textbf{Pilot} & \textbf{Pre-flight} & \textbf{Takeoff} & \textbf{Landing} & \textbf{In-flight} & \textbf{Post-flight} \\ \hline
Alpha   & 0.36043 & 0.10963  & -0.98       & -0.277  & -0.58684 \\ \hline
Bravo   & 0.38041 & -0.45515 & 0.049834  & -0.057391 & -0.20367 \\ \hline
Charlie & -0.74817 & 0.87375 & -1        & -0.046462 & -0.66847 \\ \hline
Delta   & 0.04997  & 0.44186 & -0.36667   & -0.081329 & 0.26031 \\ \hline
\end{tabular}
}
\label{tab:TempChange}
\end{center}
\end{table}

\noindent{\bf In-flight:}  Pilot Alpha, Bravo, Charlie, and Delta shows a small dip in HR below the baseline during the flight. This decrease could reflect a more settled or controlled physiological state during routine operations, potentially influenced by greater familiarity or comfort with standard in-flight tasks.

\noindent{\bf Post-flight:} Pilot Alpha, Pilot Bravo, and Pilot Charlie show HR below baseline after the flight, which could indicate relaxation post-flight. Pilot Delta shows a positive value post-flight, suggesting that they became more alert or energized after landing.

\subsubsection{Perceived Stress and Fatigue}
Pre- and post-flight Visual Analog Scale (VAS) responses were used to quantify subjective stress and fatigue levels for each participant. Across all pilots, fatigue scores increased after the flight, indicating a measurable rise in perceived tiredness following the flight session. In contrast, stress scores remained relatively consistent before and after the flight, showing minimal variation among participants. This pattern was observed consistently across all valid pilot sessions, suggesting that while pilots reported greater fatigue post-flight, their perceived stress levels showed only small decreases and did not change in any substantial way.

\subsubsection{Karolinska Sleep Questionnaire}
The sleep questionnaire results were analyzed to assess whether variations in sleep quality or non-restorative sleep could have influenced the pilots’ stress or fatigue responses. Overall, the mean scores across all participants were within a good range, indicating generally satisfactory sleep patterns as well as overall healthy sleep habits. This suggests that, for this sample, long-term sleep behavior was unlikely to be a major confounding factor in the observed responses. However, because the questionnaire reflects general sleep habits rather than specific sleep duration or quality the night before each flight, it cannot determine whether short-term sleep variability contributed to the physiological or self-reported measures collected during the study.

\subsubsection{Focused Analysis- Pilot Charlie}
In order to better understand physiological reactions during flight, a thorough case study was conducted using Pilot Charlie's data after the trends were reviewed. This subject was selected due to well-defined flight phases and obvious physiological differences. The full heart rate (HR) data recording period was analyzed, with particular attention paid to the in-flight window, which lasts roughly 20 to 88 minutes and corresponds to the actual flight time. Important in-flight events, like landing sequences and simulated engine failures, were timed using flight logs and follow-up discussions with the pilot. Electrodermal activity (EDA) was concurrently measured to see if these events were associated with increases in sympathetic arousal. 
This pilot-specific analysis provided a clearer understanding of how acute in-flight stressors manifest in physiological signals within a single participant, offering a more detailed complement to the group-level trends.

\begin{figure}[H]
    \centering
    \includegraphics[width=\columnwidth]{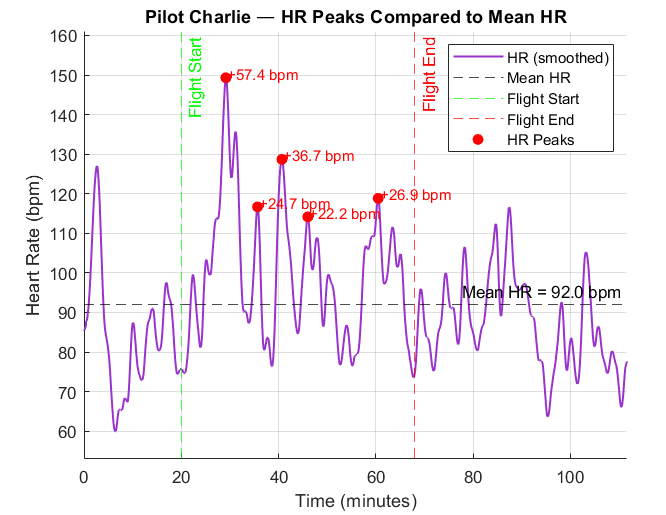}
    \caption{Pilot Charlie’s heart rate profile with identified peaks relative to the mean HR.}
\end{figure}

\begin{table}
\centering
\caption{Identified HR peaks during Pilot Charlie’s flight}
\label{tab:charlieHRpeaks}
\resizebox{\columnwidth}{!}{
\begin{tabular}{c c c c}
\hline
\textbf{Time from start (min)} & \textbf{Time from flight (min)} & \textbf{Heart Rate (bpm)} & \textbf{$\Delta$HR from Mean (bpm)} \\
\hline
29.167 & 9.167 & 149.31 & +57.36 \\
35.683 & 15.683 & 116.67 & +24.72 \\
40.683 & 20.683 & 128.69 & +36.73 \\
46.000 & 26.000 & 114.16 & +22.21 \\
60.467 & 42.467 & 118.83 & +26.88 \\
\hline
\end{tabular}
}
\end{table}

The participants' and instructors' descriptions of significant in-flight events closely correspond with the recorded HR peaks. The flight happened at 1:00 PM amid warm ambient conditions (85°F), which the pilot rated as uncomfortable. Three simulated engine failures, i.e., during the takeoff roll, at 400 feet above ground level, and on the downwind leg, occurred within a brief 2-3 minute timeframe during the class. The characteristic HR spikes displayed in Table~\ref{tab:charlieHRpeaks}, which represent acute, event-driven stress reactions, are indicative of these high-demand situations. 

Notably, the earliest and most pronounced HR spike correlates with the initial simulated engine failure during the takeoff roll, while the following peaks align with later engine-failure simulations and landing procedures. HR regularly returns to baseline values between these stressful situations, indicating a rapid physiological recovery. Temperature and electrodermal activity (EDA), two other physiological indicators, gradually rise during the flight, confirming that heart rate was the most accurate and quick marker of transient stress. 

These findings demonstrate that acute stressors during flight, particularly abrupt engine failures, cause abrupt, transient increases in heart rate that are not matched by other physiological pathways. This lends credence to the idea that heart rate is a valid indicator of short-term pilot stress in any flight situation and can react nearly instantly to acute psychological and physical demands.

\subsection{Multimodal Analysis}
A more comprehensive understanding of how pilots' internal states change throughout flying is revealed when the physiological data are analyzed together. Starting early in each session, the temperature rises steadily and consistently, reflecting both general physiological stimulation and exposure to external heat. Electrodermal activity (EDA), on the other hand, typically increases more gradually and peaks at the end of the flight, coinciding with times of increased workload and task involvement. 

Between acceleration (ACC) and EDA, a discernible pattern emerges. EDA increases in proportion to times when ACC increases, such as during post-flight aircraft handling. EDA is a potential indication of combined physical stress since this pattern supports the notion that physical effort and involvement contribute to sympathetic arousal.

These physiological alterations are further linked to subjective experience through self-reported measurements. On the PSS scale, pilot tiredness levels rose from pre-flight to post-flight, in line with the overall higher trends in temperature and EDA. This agreement implies that these physiological signs could be useful indicators of weariness during flight.

The behavior of heart rate (HR) differs from that of the other signals. HR typically increases prior to departure and progressively decreases as the session comes to a finish. This is consistent with the PSS stress trend, which exhibits either a little drop or minimal change following flight. Together, these findings show that HR may be connected to anticipatory stress or pre-flight activation, while EDA and temperature better reflect the in-flight accumulation of tiredness.

\section{Discussion}

The combined monitoring of electrodermal activity (EDA), skin temperature, and heart rate (HR) offers insight into the physiological and psychological states experienced by the pilots during flight. EDA and temperature exhibited progressive, consistent increases across all subjects, revealing larger patterns of arousal and sympathetic activity. These patterns are probably related to the combined impacts of effort, exposure to external heat, and general flight-induced weariness, placing both metrics as markers of total stress and fatigue accumulation over the course of the trip~\cite{nybo2014performance}.

On the other hand, HR offered a quicker indicator of physiological reaction. HR is sensitive to short-term shifts in workload or emotional intensity, as seen by its rapid variations during high-demand or high-focus flight phases. Because of this, HR is a helpful measure for detecting acute stress responses during particular events as opposed to general trends~\cite{fauquet2016heart}. Pilots Charlie and Delta's post-flight EDA and acceleration (ACC) peaks further indicate that their physiological reactions following shutdown were influenced by physical exertion because they both had to manually adjust their aircraft on the ramp.

When taken as a whole, these metrics provide a more comprehensive picture of pilot stress: HR shows instantaneous physiological responses to acute stressors within certain flight phases, while EDA and temperature record the steady increase in arousal over the whole flight.

\subsection{Limitations}
This study's limited sample size of four subjects is its main drawback. It is impossible to make trustworthy inferences or extrapolate the results to the larger pilot population with such a small sample size. Additionally, it was not possible to sufficiently account for variations in training progress, procedural familiarity, and individual coping mechanisms. Even slight variations in competence or comfort can affect stress and fatigue reactions, even if the study's participants were assumed to have similar levels of expertise. Because of the small sample size, past research in other transportation domains, such as automobile studies, has demonstrated that more experienced operators commonly exhibit lower workload and fatigue patterns that could not be adequately addressed here~\cite{Li2016Fatigue}.  
%
%
Further, the sleep questionnaire recorded general sleep habits rather than sleep quality on the night before each travel. As a result, short-term sleep variability, which might influence stress, alertness, and physiological responses, could not be evaluated~\cite{Dawson2005ManagingFatigue}.

\subsection{Future Work}
To enable meaningful statistical analyses and better capture variability across experience levels, future work will involve a bigger datasets with more diverse participant pool, such as NetHealth physiological and behavioral dataset~\cite{vhaduri2021predicting}. A broader context for interpreting physiological responses would be provided by incorporating additional psychological and behavioral metrics, such as reaction-time tests and other signs of cognitive effort, along with geo-temporal pattern analysis~\cite{vhaduri2021deriving}.

Direct environmental data, such as noise levels, whole-body vibration, and temperature stress, will also be included in future research. Collecting these measures will help distinguish between physiological responses driven by workload and those influenced by environmental conditions. Future research will also concentrate on creating generic and customized stress and fatigue estimation with advanced models, such as quantum models~\cite{vhaduri2026we} or large language models (LLMs)~\cite{gammon2026many}, as well as protecting data privacy~\cite{vhaduri2022predicting}.

Pilots would not have to remember takeoff times, maneuver sequences, and other stressful situations if automated event-recording systems or cockpit-based time markers were included and embedded this environmental/information knowledge into modeling~\cite{vhaduri2023environment}. Aligning physiological data with flight events would become far more accurate as a result.

\subsection{Practical Implications}
There are various useful advantages to measuring physiological reactions during flight. First, it helps pilots become more conscious of their own levels of stress and exhaustion, which promotes self-monitoring and safer decision-making. Second, these metrics could be used by flying schools and training programs to determine when specific assistance or intervention may be required to enhance pilot wellbeing. Lastly, by providing data-driven insight into how stress develops and appears during actual flight operations, these findings support larger organizational initiatives in fatigue-risk management.

\section*{Acknowledgement} 
ChatGPT has been used to enhance part of the writing.


\bibliographystyle{IEEEtran}
\bibliography{reference}
\end{document}